\documentclass[preprint,aps,amsfonts,nofootinbib,superscriptaddress,preprintnumbers]{revtex4}

\usepackage{amscd}
\usepackage{amsmath}
\usepackage{graphicx,subfigure}
\usepackage{comment}

\input xy
\xyoption{all}

\newcommand{\beq}{\begin{equation}}
\newcommand{\eeq}{\end{equation}}
\newcommand{\bea}{\begin{eqnarray}}
\newcommand{\eea}{\end{eqnarray}}

\newcommand{\nn}{\nonumber}
\newcommand{\benn}{\begin{displaymath}}
\newcommand{\eenn}{\end{displaymath}}

\newcommand{\tr}{{\rm tr}}
\newcommand{\str}{{\rm str\ }}

\newcommand{\D}{{\mathcal D}}

\renewcommand{\l}{\left}

\newcommand{\x}{{\mathbf{x}}}

\def\slashchar#1{\ensuremath{                               %
   \setbox0=\hbox{${}#1{}$}       
   \dimen0=\wd0                                 
   \setbox1=\hbox{/} \dimen1=\wd1               
   \ifdim\dimen0>\dimen1                        
      \rlap{\hbox to \dimen0{\hfil/\hfil}}      
      {}#1{}                                    
   \else                                        
      \rlap{\hbox to \dimen1{\hfil${}#1{}$\hfil}}   
      /                                         
   \fi}}                                        %
\def\simge{
    \mathrel{\rlap{\raise 0.511ex
        \hbox{$>$}}{\lower 0.511ex \hbox{$\sim$}}}}
\def\simle{
    \mathrel{\rlap{\raise 0.511ex
        \hbox{$<$}}{\lower 0.511ex \hbox{$\sim$}}}}

\begin{document}

\def\a{{\alpha}}
\def\b{{\beta}}
\def\d{{\delta}}
\def\D{{\Delta}}
\def\e{{\epsilon}}
\def\g{{\gamma}}
\def\G{{\Gamma}}
\def\k{{\kappa}}
\def\l{{\lambda}}
\def\L{{\Lambda}}
\def\m{{\mu}}
\def\n{{\nu}}
\def\w{{\omega}}
\def\O{{\Omega}}
\def\S{{\Sigma}}
\def\s{{\sigma}}
\def\t{{\tau}}
\def\th{{\theta}}
\def\x{{\xi}}

\def\ol#1{{\overline{#1}}}

\def\Dslash{D\hskip-0.65em /}
\def\dslash{{\partial\hskip-0.5em /}}
\def\vslash{{\rlap \slash v}}
\def\qbar{{\overline q}}

\def\CPT{{$\chi$PT}}
\def\QCPT{{Q$\chi$PT}}
\def\PQCPT{{PQ$\chi$PT}}
\def\tr{\text{tr}}
\def\str{\text{str}}
\def\diag{\text{diag}}
\def\order{{\mathcal O}}
\def\vit{{\it v}}
\def\vD{\vit\cdot D}
\def\am{\alpha_M}
\def\bm{\beta_M}
\def\gm{\gamma_M}
\def\smb{\sigma_M}
\def\smt{\overline{\sigma}_M}
\def\tb{{\tilde b}}

\def\c#1{{\mathcal #1}}

\def\Bbar{\overline{B}}
\def\Tbar{\overline{T}}
\def\cBbar{\overline{\cal B}}
\def\cTbar{\overline{\cal T}}
\def\pq{(PQ)}

\def\eqref#1{{(\ref{#1})}}

\preprint{NT@UW 05-15}
\preprint{LBNL-59327}

\title{Finite volume corrections to $\pi\pi$ scattering} 
\author{Paulo F.~Bedaque}
\email[]{bedaque@umd.edu}
\affiliation{Lawrence-Berkeley Laboratory, Berkeley, CA 94720, USA}
\affiliation{University of Maryland, College Park, MD 20742, USA} 
\author{Ikuro Sato}
\email[]{isato@lbl.gov}
\affiliation{Lawrence-Berkeley Laboratory, Berkeley, CA 94720, USA}
\author{Andr\'e Walker-Loud} 
\email[]{walkloud@u.washington.edu}
\affiliation{University of Washington, Box 351560, Seattle, WA 98195-1560, USA}

\begin{abstract}
 Lattice QCD studies of hadron-hadron interactions are performed by computing the energy levels of the system in a finite box. The shifts in energy levels proportional to inverse powers of the volume are related to scattering parameters in a model independent way.  In addition, there are non-universal exponentially suppressed corrections that distort this relation. These terms are proportional to $e^{-m_\pi L}$ and become relevant as the chiral limit is approached. In this paper we report on a one-loop chiral perturbation theory calculation of the leading exponential corrections in the case of $I=2$ $\pi\pi$ scattering near threshold.
\end{abstract}
\maketitle


%
%
%
%
%
%
%
%
%
%
%
%
\section{Introduction}
Except in the case of infinitely heavy baryons, where an adiabatic potential can be defined, the interaction between two hadrons is studied with lattice QCD by numerically calculating energy levels of the system in a finite box. This is because in the infinite volume limit and away from kinematical thresholds, the two-hadron Euclidean correlator gives no information about the Minkowski space amplitude~\cite{maiani-testa}. The  alternative is to consider the system in a finite box, as is the case with numerical calculations anyway. The energy levels of a system composed of two hadrons are not simply the sum of the energies carried by each hadron, but there is an additional (usually small) shift that arises due to the interaction between them. The smaller the box, the larger the shift in energy levels.  This volume dependence is inversely proportional to the volume and furthermore, there is a relation between the energy level shifts and the scattering phase shifts~\cite{hamber_et_al,luscher_1,luscher_2}. This relation, valid for energies below the first inelastic threshold is a consequence of unitarity and is thus model 
independent.%
\footnote{By model independent relation we mean  a relation valid whether one is considering QCD or some other theory, as long as this theory obeys unitarity, locality, \textit{etc}.}
In addition to this power law shift in the energy levels, there are exponentially suppressed corrections which are {\it not} model independent and are the analogue of the exponentially suppressed corrections to the mass, decay constants, \textit{etc}., in the single-hadron sector~%
\cite{Luscher:1985dn,gasser_volume,Colangelo:2003hf,Beane:2004tw,colangelo_2loops}. These exponential volume effects arise because the off-shell propagation of intermediate states is altered by the presence of the finite box, which allows them, for instance, to ``wrap around" the lattice.  As such, these effects are dominated by the lightest particle, the pion in QCD, and are proportional to $e^{-m_\pi L}$ with $m_\pi$ the pion mass and $L$ the linear dimension of the box. For simulations done with small enough quark masses such that the pions are within the chiral regime, these soft pion effects can be computed using the chiral perturbation theory (\CPT)~\cite{Gasser:1983yg,Gasser:1984gg}.
The $\pi\pi$ scattering phase shifts have been computed using lattice QCD following the universal finite volume method mentioned above~\cite{Sharpe:1992pp,Gupta:1993rn,Kuramashi:1993ka,Fukugita:1994na,Fukugita:1994ve,Aoki:1999pt,Alford:2000mm,Liu:2001zp,Aoki:2001hc,Aoki:2002in,Aoki:2002ny,Juge:2003mr,Yamazaki:2004qb,Du:2004ib,Gattringer:2004wr,Aoki:2005uf,Beane:2005rj}. As the chiral limit 
is approached~\cite{Beane:2005rj} and more precise calculations appear, these exponentially suppressed corrections will need to be understood. 

Our goal in this paper is first to show the modification of 
the universal scattering formula for a hadron-hadron system in a box
due to the exponentially suppressed finite volume corrections.
Second, we compute the dominant exponential volume dependence explicitly
for the two-pion system in $I=2$ channel near threshold by use of the 
leading loop-order two-flavor \CPT.

%
%
%
\section{Finite Volume $\pi\pi$ Scattering}

\subsection{Power law and exponential volume dependence}

As discussed above there  are two types of volume dependence of the energy levels of two hadrons in a box: power law (proportional to $1/L^3$) and exponential (proportional to $e^{-m_\pi L}$). The first is exploited by the finite volume method to extract information about scattering parameters~\cite{hamber_et_al,luscher_1,luscher_2}. The second, usually numerically smaller, appears as a correction to the relation between energy levels in a box and scattering parameters.  In order to compute the exponentially suppressed terms we need to separate them from the larger power law contribution.  

\begin{figure}[b]\label{fig:feynman}
\begin{tabular}{ccccc}
	\includegraphics[width=0.28\textwidth]{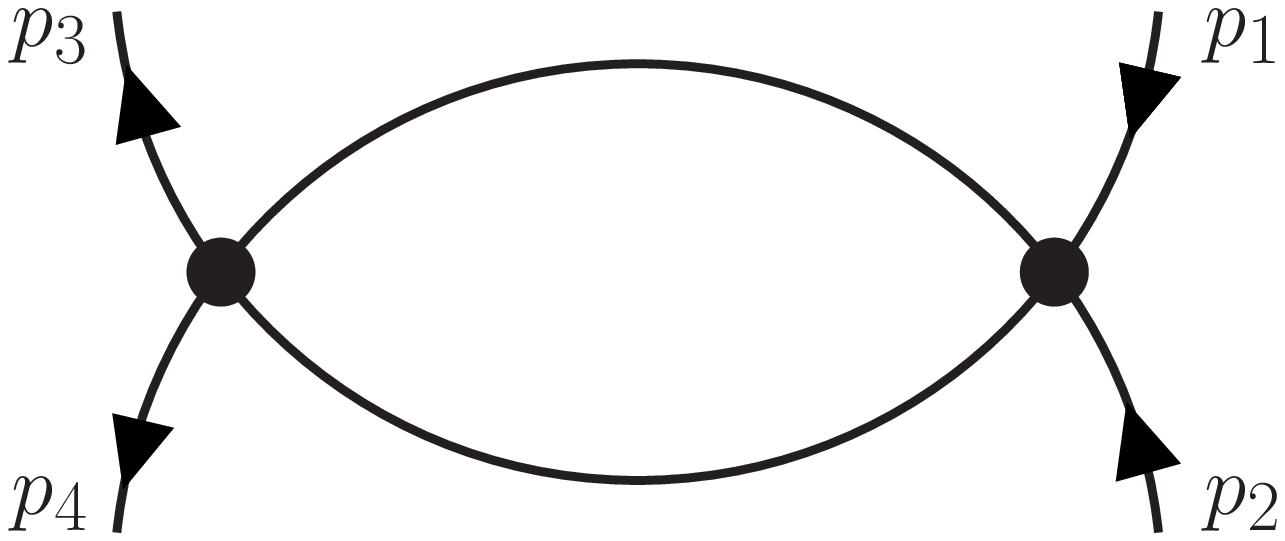} & $\;\;\;\;$ & \includegraphics[width=0.28\textwidth]{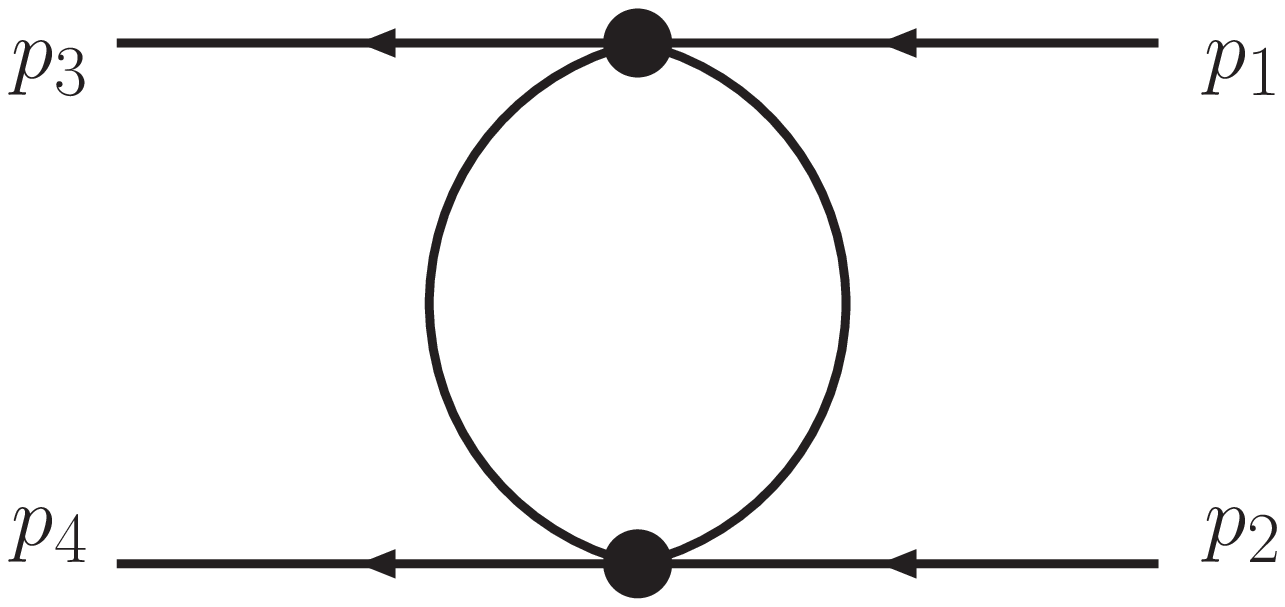} & $\;\;\;\;$ & \includegraphics[width=0.28\textwidth]{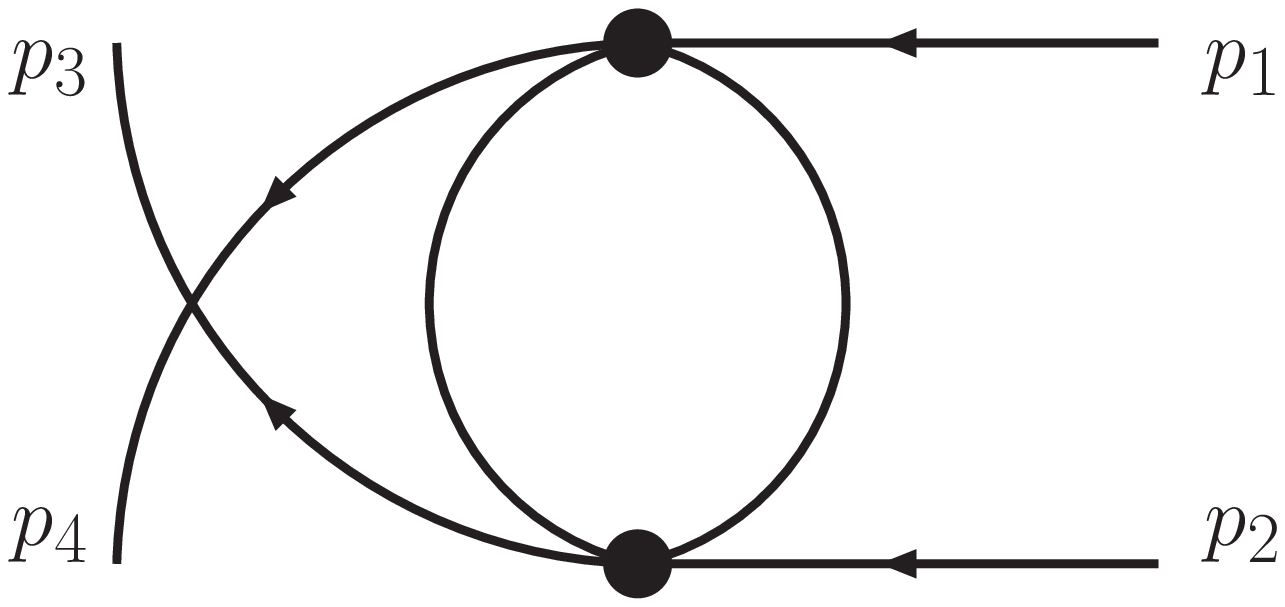} \\
	(a) & $\;\;$ & (b) & $\;\;$ & (c) \\
\end{tabular}
\vspace{2mm}
\begin{tabular}{cccc}
	\includegraphics[width=0.19\textwidth]{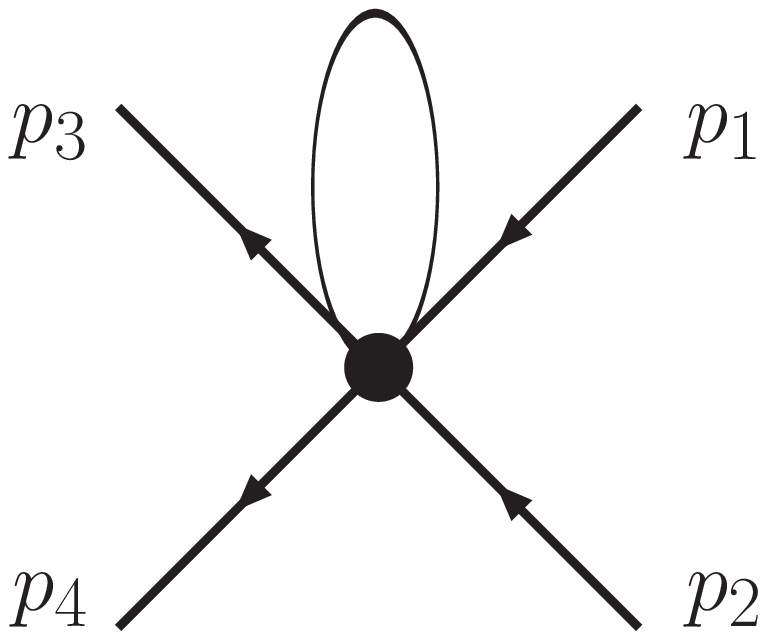} & $\;\;\;\;\;\;$ & \includegraphics[width=0.19\textwidth]{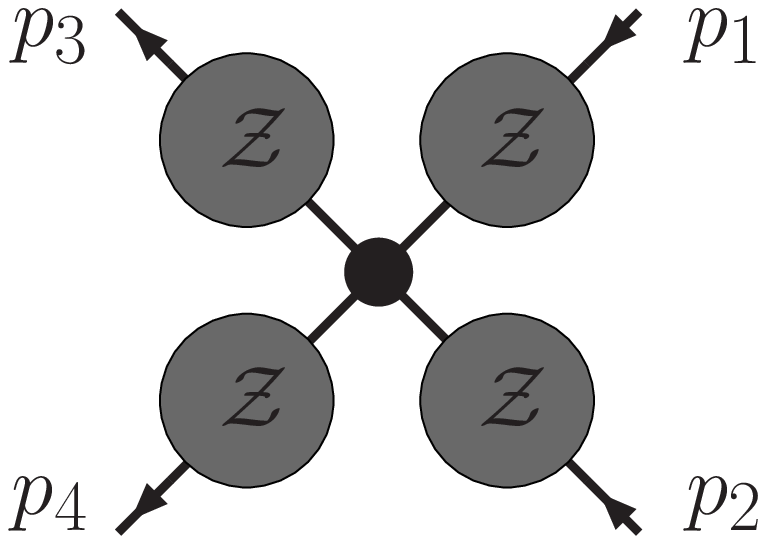}  \\
	(d) & $\;\;\;\;\;\;\;\;\;\;\;\;\;\;$ & (e)  \\
	\end{tabular}
	\caption{The one-loop diagrams which contributing to the $\pi\pi$ scattering amplitude.  Only the $s$-channel diagram, (a), contributes to the power-law volume dependence.  Diagrams (b) and (c) are the $t$-, and $u$-channel diagrams, respectively, while diagram (e) represents wavefunction renormalization.  All these diagrams contribute to the exponential volume dependence.}
\label{fig:one-loop}
\end{figure}

Figure~\ref{fig:one-loop} shows all $\pi\pi$ scattering diagrams which contribute at one-loop order. As we will discuss in more detail in the next section, the power law corrections arise only from 
\nobreak{$s$-channel} diagrams as shown in Fig.~\ref{fig:one-loop}~(a), where the intermediate particles can be on-shell, and thus propagate far and ``feel'' the finiteness of the box. In all other diagrams the intermediate particles are very off-shell, cannot propagate farther than a distance of order $1/m_\pi$ and therefore have only small, exponentially suppressed sensitivity to the size of the box. 

Let us now discuss the general form of finite volume corrections.
Consider first the pion propagator at finite volume which is a function of the spatial momentum $\vec{k}=2\pi \vec{n}/L$ and the energy $E$. It will have poles for values of $E$ corresponding to the values of the energy of a pion in the box. In particular, for $\vec{k}=0$ the pole will be at $m_\pi(L)$ (the ``finite volume mass"), differing from the (infinite volume) mass of the pion $m_\pi$ by an exponentially small quantity proportional to $(m_\pi^2 / (4\pi f)^2)\, e^{-m_\pi L} / L\sqrt{mL}$~%
\cite{gasser_volume,Colangelo:2003hf,colangelo_2loops}. 
The extra suppression factor $m_\pi^2/(4\pi f_\pi)^2$ is due to the fact that only loop diagrams contribute to the finite volume corrections.

The volume corrections for systems with more than one hadron are more subtle. The reason is that there are two kinds of volume corrections to the energy levels: a power law one described by the L\"{u}scher formula and the exponentially suppressed ones. To understand how to separate them let us first look at the infinite-volume, S-wave scattering amplitude, $T(s)$, with on-shell external pions. It is given at one loop by
\footnote{We are considering the s-wave projected amplitude and disregarding the mixing with higher partial waves induced by the breaking of rotational symmetry.}:
\bea\label{eq:Tgeneral}
T(s) &\simeq& T^{(0)}(s) +T^{(1)} _{t,u}(s) + T^{(1)}_{s,R} (s) + i T^{(1)}_{s,I}(s)\nn\\
&\simeq& \frac{(T^{(0)}(s))^2}{T^{(0)}(s) -T^{(1)}_{t,u}(s) - T^{(1)}_{s,R} (s) - i T^{(1)}_{s,I}(s)},
\eea
where $T^{(n)}$ is $n$-th loop contribution, the $s$-channel contribution is separated into its real part, $T^{(1)}_{s,R}(s)$, and imaginary part, $T^{(1)}_{s,I}(s)$, and all other contributions at one-loop including $t$- and $u$-channels are denoted by $T^{(1)}_{t,u}(s)$. The imaginary part, $T^{(1)}_{s,I}(s)$, comes from picking in the loop integration, the particle poles in both pion propagators, such that both the loop pions are on-shell. The loop integral is then proportional to the phase space volume and is given by 
\begin{equation}
T^{(1)}_{s,I}(s) = {(T^{(0)}(s))^2\over 32\pi\sqrt{s}} \sqrt{s-4m_\pi^2}. 
\end{equation}
The fact that the imaginary part is determined by the tree level amplitude is a consequence of the optical theorem.

It is useful to define the $K$-matrix~\cite{taylor}, which at one-loop is given by:
\beq\label{eq:K}
K(s) \simeq T^{(0)}(s)  +T^{(1)} _{t,u}(s) + T^{(1)}_{s,R} (s)
\simeq  \frac{(T^{(0)}(s))^2}{T^{(0)}(s) -T^{(1)} _{t,u}(s) - T^{(1)}_{s,R} (s)}.
\eeq 
Since the scattering amplitude can be written in terms of the phase shift, $\delta(s)$, as
\beq
T(s) = \frac{32\pi\sqrt{s}}{\sqrt{s-4m_\pi^2}}\frac{1}{\cot\delta(s)-i}
= \frac{32\pi\sqrt{s}}{\sqrt{s-4m_\pi^2}}\frac{e^{i\delta(s)}-1}{2i},
\eeq 
the relation between the $K$-matrix and the phase shift is then given by
\beq\label{eq:kcot}
\frac{1}{K(s)} = \frac{1}{32\pi}\sqrt{\frac{s-4m_\pi^2}{s}} \cot\delta(s).
\eeq  

Now we look at the finite volume amplitude $\mathcal{T}(s)$\footnote{By finite volume scattering amplitude we mean the amputated four-point correlator since, of course, there is no scattering at finite volume.}. It is computed in the same way as the infinite volume amplitude, except the loop integrals are substituted by sums over the  momenta allowed in a finite box.
The important point to keep in mind is that sums where the summand is regular are, at large enough $L$, well approximated by the analogous integral, up to exponentially small terms. If the summand, however, contains a singularity, power law dependence on the volume arises.
As mentioned before, only the kinematics of the $s$-channel diagram allows for both of the intermediate pions to be on-shell simultaneously. This implies that the summand in the sum over the loop momentum contains a singularity and leads to power law volume corrections. For the remaining diagrams no singularities are present and only exponentially suppressed corrections can arise.
As it will be shown explicitly below, the finite volume amplitude then has the following form: the tree term remains the same, the $t$-, $u$- and the real part of the $s$-channels pick only exponential corrections but the imaginary part turns into the term with power law $L$-dependence (and is real at finite $L$):
\bea\label{eq:finiteT}
\mathcal{T}(s) &\simeq& T^{(0)}(s) + T^{(1)} _{t,u}(s) + T^{(1)}_{s,R} (s) +\Delta T_{\rm exp}^{(1)}(s) + \frac{(T^{(0)}(s))^2}{16\pi^2 L\sqrt{s}}\mathcal{S}\left( \frac{k^2L^2}{4\pi^2} \right) \nn\\
&\simeq& \frac{(T^{(0)}(s))^2}{T^{(0)}(s)-  T^{(1)} _{t,u}(s) - T^{(1)}_{s,R} (s) -\Delta T_{\rm exp}^{(1)}(s) - \frac{(T^{(0)}(s))^2}{16\pi^2 L\sqrt{s}}\mathcal{S}\left( \frac{k^2L^2}{4\pi^2} \right) }.
\eea 
where $s=4(k^2+m_\pi^2)$, $\Delta T_{\rm exp}^{(1)}(s)$ is the finite volume correction to $T^{(1)} _{t,u}(s) + T^{(1)}_{s,R} (s)$, and $\mathcal{S}$ is a universal (independent of the interaction) function of 
$s$~\cite{Beane:2003yx,Beane:2003da}:
\beq
\mathcal{S} \left( \frac{k^2L^2}{4\pi^2} \right) 
	= 4\pi^2 L \left[ \frac{1}{ L^3}\sum_{\vec{q}=\frac{2\pi \vec{n}}{L}} -\int \frac{d^3q}{(2\pi)^3} \right] \frac{1}{\vec{q}^2-k^2}
	= \lim_{\Lambda_n\rightarrow\infty}\sum_{|\vec{n}|<\Lambda_n}\frac{1}{\vec{n}^2-\frac{k^2L^2}{4\pi^2}} -4\pi\Lambda_n.
\eeq
Combining Eqs.(\ref{eq:Tgeneral},\ref{eq:K},\ref{eq:kcot},\ref{eq:finiteT}) we have
\bea
\mathcal{T}(s) &\simeq& \frac{1}{\frac{1}{K(s)} - \frac{\Delta T_{\rm exp}^{(1)}(s)}{(T^{(0)}(s))^2} - \frac{1}{16\pi^2 L\sqrt{s}}\mathcal{S}\left(\frac{(s-4m_\pi^2)L^2}{16\pi^2}\right)}\nn\\
&=& \frac{16\pi \sqrt{s}}{k\cot\delta(s)- 16\pi\sqrt{s}\frac{\Delta T_{\rm exp}^{(1)}(s)}{(T^{(0)}(s))^2} -\frac{1}{\pi L}\mathcal{S}\left(\frac{(s-4m_\pi^2)L^2}{16\pi^2}\right) }.
\eea 
The energy of the states in the box are determined by the location of the poles of the finite volume amplitude  determined by the solution of
\beq\label{eq:FVluscher}
k\cot\delta(s)- 16\pi\sqrt{s}\frac{\Delta T_{\rm exp}^{(1)}(s)}{(T^{(0)}(s))^2} = \frac{1}{\pi L}\mathcal{S}\left(\frac{(s-4m_\pi^2)L^2}{16\pi^2}\right).
\eeq 
We recognize in Eq.~(\ref{eq:FVluscher}) the familiar form of the L\"{u}scher relation modified by the
finite volume correction: the quantity $- 16\pi\sqrt{s}\frac{\Delta T_{\rm exp}^{(1)}(s)}{(T^{(0)}(s))^2} $ is the sought-after (exponentially small) correction to $k\cot\delta(s)$. 
In this work, we will focus on the 2-pion correlator near threshold in the center of mass frame, for which the infinite volume energy is given by $\sqrt{s}=2m_\pi$.
The solution of Eq.~(\ref{eq:FVluscher}), $s^*$, will be away from threshold by an amount given by $\sqrt{s^*}-2m_\pi$~$\approx \sqrt{s^*}-2m_\pi(L)$~$\sim 1 / f_\pi^2L^3$. Therefore, for $s\approx s^*$ the correction term, 
$\D(k \cot\delta(s)) = - 16\pi\sqrt{s}\frac{\Delta T_{\rm exp}^{(1)}(s)}{(T^{(0)}(s))^2} $,  can be approximated by $- 32\pi m_\pi\frac{\Delta T_{\rm exp}^{(1)}(4m_\pi^2)}{(T^{(0)}(4m_\pi^2))^2} $, the difference being suppressed by $\sim 1/L^3$.

It is customary to expand Eq.~(\ref{eq:FVluscher}) in  powers of $k^2 \sim 1/L^3$. Up to  the first three orders of this expansion (near threshold), $k\cot\delta$ can be approximated by the inverse scattering length, $1/a$, resulting in 
\beq\label{eq:luscher_approx}
	\sqrt{s^*}-2 m_\pi = \frac{4\pi a}{m_\pi L^3}
				\left( 1 + c_1 \frac{a}{L} + c_2\left( \frac{a}{L}\right)^2 +\cdots\right),
\eeq 
where $c_{1,2}$ are known numerical factors. A generalization of this formula  including the exponentially suppressed corrections, however, is not useful. The error in using Eq.~(\ref{eq:luscher_approx}) instead of Eq.~(\ref{eq:FVluscher}), that is, the error in the extrapolation from $s^*$ to $4m_\pi^2$, is of order $1/L^3$, which is parametrically larger than the exponential corrections we are interested in.  Numerically,  it may be the case that, for a set of simulation parameters,  the exponential term is larger than the $1/L^3$ terms. But the analogue of  Eq.~(\ref{eq:luscher_approx}) one would obtain by formally counting, for instance $e^{-m_\pi L} \sim 1/L^2$, would involve the effective range in addition to the scattering length. In any case, it is unclear {\it a priori} that a set of simulation parameters exist where this kind of expansion is useful and we will not pursue this line of thought in this paper.
We now will compute the exponential corrections to $k\cot \delta(s)$.
%


%
%
%
\subsection{The $\pi\pi$ scattering amplitude}
\label{sec:fvT}

The  $\pi\pi$ finite volume correlator in the $I=2$ isospin channel, for arbitrary momentum (in the chiral regime) is given by
\begin{align}\label{eq:Ttensor}
	\c{T}_2 
	&= -\frac{2}{f^2} \bigg\{
			\frac{(3s -2m^2-\sum_{i=1}^4 p_i^2)}{3} 
		+\Big[ \frac{10s}{9f ^2} -\frac{6m ^2}{ f ^2}
			\Big]\, i\c{I} 
		+\frac{4}{9 f ^2} p_s^\mu p_s^\nu\, i\c{J}_{\mu\nu} ( p_s) \nonumber\\
	&\qquad\qquad
		+ \Big[ \frac{4}{9 f ^2} p_t^\mu p_t^\nu 
			+\frac{2}{9 f ^2} \Big( p_t +3 (p_1 +p_3) \Big)^\mu 
				\Big( p_t +3 (p_2 +p_4) \Big)^\nu \Big]\, i\c{J}_{\mu\nu}( p_t) \nonumber\\
	&\qquad\qquad
		+ \Big[ \frac{4}{9 f ^2} p_u^\mu p_u^\nu 
			+\frac{2}{9 f ^2} \Big( p_u +3 (p_1 +p_4) \Big)^\mu 
				\Big( p_u +3 (p_2 +p_3) \Big)^\nu \Big]\, i\c{J}_{\mu\nu}( p_u) \nonumber\\
	&\qquad\qquad
		+\frac{4(s-3m ^2)^2}{9 f ^2}\, i\c{J}( p_s) 
		+\Big[ \frac{m ^4}{f ^2} -\frac{4 t m ^2}{3 f ^2} +\frac{2 t^2}{3 f ^2}
			\Big]\, i\c{J}( p_t) \nonumber\\
	&\qquad\qquad
		+\Big[ \frac{m ^4}{f ^2} -\frac{4 u m ^2}{3 f ^2} +\frac{2 u^2}{3 f ^2}
			\Big]\, i\c{J}( p_u) 
		-\frac{8(s-3m ^2)}{9f ^2}\, p_s^\mu\, i\c{J}_\mu( p_s) 
		+\frac{4 (m ^2-t)}{3 f ^2} p_t^\mu \, i\c{J}_\mu( p_t) \nonumber\\
	&\qquad\qquad
		+\frac{4 (m ^2-u)}{3 f ^2} p_u^\mu \, i\c{J}_\mu( p_u) 
		-\frac{4 \ell_1}{f ^2} \Big[ (t-2m^2)^2 +(u- 2m^2)^2 \Big] \nonumber\\
	&\qquad\qquad
		-\frac{2 \ell_2}{f ^2} \Big[ 2(s-2m^2)^2 +(t-2m^2)^2 +(u-2m^2)^2 \Big] 
		-\frac{32}{3}\ell_3 \frac{m ^4}{f ^2}
		\bigg\}.
\end{align}
In the above expression, $m$ is the (volume independent) tree level pion mass, $f\approx 132$~MeV is the (volume independent) tree level decay constant and the $\ell_i$'s are the Gasser-Leutwyler coefficients of counter terms appearing in the chiral Lagrangian at next-to-leading order (NLO)~\cite{Gasser:1983yg}.  The loop integrals/sums are given by $\c{I}$, $\c{J}(P)$, $\c{J}_\mu(P)$, and $\c{J}_{\mu\nu}(P)$ and will be defined below shortly.  The external momenta $p_i,\;i=1\cdots4$ are described in Fig.~\ref{fig:one-loop}, and the Mandelstam variables are employed: $s=p_s^2$ with $p_s=p_1+p_2$, $t=p_t^2$ with $p_t = p_1-p_3$, and $u=p_u^2$ with $p_u=p_1-p_4$.  In the above equation, the first term is the leading-order (LO) tree level contribution, and the remaining terms come from the one-loop diagrams shown in Fig.~\ref{fig:one-loop} and tree level diagrams with vertices of the $\c{O}(p^4)$ Lagrangian~%
\cite{Gasser:1983yg,Gasser:1984gg}. In the NLO contributions, we have approximated $p_i^2 = m_\pi^2$, as the corrections to this at finite volume are beyond the order we are working.
%
%
In the one-loop terms, we have expressed all contributions in terms of the bare pion mass as the difference is higher order in the chiral expansion.
One can, of course, choose to express the scattering amplitude in terms of the ``lattice quantities'' such as $m_\pi(L)$ and $f_\pi(L)$, which are measured directly from Euclidean correlation functions by lattice simulations.  Converting the bare quantities into lattice quantities involves additional tadpole loops, which will affect the form of $\D \c{T}_{\rm exp}^{(1)}$, the finite volume corrections to the two-pion amplitude.  The exponential volume dependence of course doesn't depend upon whether one expresses the amplitude in terms of either the bare of physical parameters, and so it is useful to use the form which is simplest, that in terms of bare parameters.  In what follows, we will be interested in the 2-pion correlator near threshold, for which the external pion momentum are given by 
$p_i \simeq (\frac{1}{2}\sqrt{s}, \vec{0})$.

%
%
%
%
\subsection{Loop integrals/sums at one-loop}

The loop integrals/sums appearing in Eq.~(\ref{eq:Ttensor}) are defined by
\begin{align}
 \mathcal{I}  &= \int \frac{dq_0}{2\pi} \frac{1}{L^3}\sum_{\vec{q} = \frac{2\pi \vec{n}}{L}} 
                         \frac{1}{q^2-m ^2}, \label{eq:intI}   \\ 
\mathcal{J}(P) &= \int \frac{dq_0}{2\pi} \frac{1}{L^3}\sum_{\vec{q} = \frac{2\pi \vec{n}}{L}} 
                         \frac{1}{q^2-m ^2}   \frac{1}{(P+q)^2-m ^2}, \label{eq:intJ} \\
\mathcal{J}_\mu(P) &= \int \frac{dq_0}{2\pi} \frac{1}{L^3}\sum_{\vec{q} = \frac{2\pi \vec{n}}{L}} 
                         \frac{q_\mu}{q^2-m ^2}   \frac{1}{(P+q)^2-m ^2}, \label{eq:intJmu}
\end{align}
and
\begin{equation}
\mathcal{J}_{\mu\nu}(P) = \int \frac{dq_0}{2\pi} \frac{1}{L^3}
\sum_{\vec{q} = \frac{2\pi \vec{n}}{L}} \frac{q_\mu q_\nu}{q^2-m ^2} \frac{1}{(P+q)^2-m ^2}. 
\label{eq:intJmunu}
\end{equation}
Note that an integral is taken along the $0^{\rm th}$ component whereas sums over discrete momenta are taken with cubic symmetry.  Finite volume effects in the loop integrals/sums in Eqs.~\eqref{eq:intI}-\eqref{eq:intJmunu} can be computed by first evaluating the $q_0$ contour integral and then using the Poisson resummation formula,
\beq\label{eq:poisson}
	\frac{1}{L^3}\sum_{\vec{q} =
		\frac{2\pi \vec{n}}{L}} f(\vec{q}) =
		\int\frac{d^3q}{(2\pi)^3}f(\vec{q})
			+\sum_{\vec{n}\neq 0, \vec{n}\in \mathbb{N}^3} \int\frac{d^3q}{(2\pi)^3} f(\vec{q}) e^{iL\vec{q}\cdot\vec{n}}.
\eeq 
The difference between the finite volume and infinite volume loop integrals/sums is given by the second term in the right-hand side of Eq.~(\ref{eq:poisson}), and is always ultraviolet finite.  If the function $f(\vec{q})$ is regular, this difference is exponentially suppressed in the large $L$ limit. Power law dependence on $L$ can however appear if $f(\vec{q})$ has a singularity, {\it i.e.}, the case when $P=p_s$.  

Let us now evaluate the \textit{difference} between the finite and infinite volume integrals/sums given in Eqs.~(\ref{eq:intI})-\eqref{eq:intJmunu}.  We shall define the difference between the finite and infinite volume integrals/sums as
\begin{equation}
	\D f \equiv f(FV) - f(\infty) =
		 \int \frac{dq_0}{2\pi} \left[ \frac{1}{L^3}\sum_{\vec{q}= \frac{2\pi \vec{n}}{L}}
				- \int \frac{d^3 q}{(2\pi)^3} \right] 
			f(\vec{q}),
\end{equation}
where it is implicit that we regulate the sum and integral in the same manner such that the UV divergences cancel.  The tadpole integral in Eq.~\eqref{eq:intI}, which contributes to $m_\pi$ and $f_\pi$ and the loop diagrams of Fig.~\ref{fig:one-loop}, has the following volume correction;
\begin{align} \label{eq:deltaI}
	i\Delta \c{I} &= \int \frac{dq_0}{2\pi} \left[ \frac{1}{L^3}\sum_{\vec{q}= \frac{2\pi \vec{n}}{L}}
					- \int \frac{d^3 q}{(2\pi)^3} \right] 
				\frac{i}{q^2-m ^2} \nonumber\\
			&= \left[\frac{1}{L^3}\sum_{\vec{q}}-\int\frac{d^3 q}{(2\pi)^3}\right]\frac{1}{2\omega_q},
		\nonumber\\
			&= \frac{m}{4\pi^2 L} \sum_{\vec{n}\neq 0} 
				\frac{1}{|\vec{n}|} K_1(|\vec{n}| m  L).
\end{align}
where $\omega_q=\sqrt{\vec{q}^{\:2}+m^2}$.  The mass and decay constant measured in lattice simulations are thus given to NLO by~\cite{Colangelo:2003hf}%
\footnote{This relation is known up to two loops~\cite{colangelo_2loops,Bijnens:2005ne}.}
\begin{align}
	m_\pi^2(L) &= m_\pi^2 \left[ 1+\frac{i \Delta\c{I}}{f_\pi^2} \right] 
			= m^2 \left[ 1+\frac{i \c{I}(L=\infty) + i \D \c{I}}{f^2} 
				+ \frac{4 \ell_3 m^2}{f^2} \right], \\
	f_\pi(L) &= f_\pi \left[ 1- \frac{2 i \D \c{I}}{f_\pi^2} \right]
			= f \left[ 1-\frac{2i\c{I}(L=\infty) + 2i \D \c{I}}{f^2}
				+\frac{2 \ell_4 m^2}{f^2}\right].
\end{align}
Using the asymptotic form of the Bessel function, one can see that for large $m L$, the volume shift of the pion mass is exponential~\cite{gasser_volume,Colangelo:2003hf},
\begin{equation}
	\frac{\D m_\pi^2}{m_\pi^2} = \frac{i\D \c{I}}{f_\pi^2} 
		= \frac{1}{2^{5/2} \pi^{3/2}} \frac{m_\pi}{L f_\pi^2} \sum_{n=|\vec{n}| \neq 0} 
			\frac{e^{-n\, m_\pi L}}{n^{3/2}} \frac{c(n)}{\sqrt{m_\pi L}}
			\left[ 1 + \frac{3}{8} \frac{1}{n\, m_\pi L} + \ldots \right],
\end{equation}
where the ellipses denote more terms in the asymptotic expansion of the Bessel function and $c(n)$ is the multiplicity factor counting the number of times $n=|\vec{n}|$ appears in the 3-dimensional sum.  Note, this sum is not over integers, but rather over the square-roots of integers.  In Table~\ref{t:c(n)}, we list the first few values of the multiplicity factors.

\begin{table}[t]
\caption{Here we list the first few multiplicity factors which arise when converting the three-dimensional sum to a scalar sum.}
\begin{tabular}{| c | c c c c c |}
\hline
$n$
 & $1$ 
 & $\sqrt{2}$ 
 & $\sqrt{3}$ 
 & $\sqrt{4}$ 
 & $\sqrt{5}$ \\
\hline
$c(n)$
 & $6$ 
 & $12$ 
 & $8$ 
 & $6$ 
 & $24$ \\
\hline
\end{tabular}
\label{t:c(n)}
\end{table}

\bigskip

Power law $L$-dependence can only occur through the integrals/sums in Eqs.~\eqref{eq:intJ}-\eqref{eq:intJmunu} when $P^2>0$. For the  center-of-mass scattering kinematics we are considering here this can only occur for $P=p_s$, since $p_s^2 = s>0$. As argued above we will only need the amplitude at threshold, \textit{i.e.}, $p_s=(2m_\pi,\vec{0})$ and $p_t=p_u=0$, except for the terms with power law $L$-dependence. Consequently, we will need only the values of $\Delta \mathcal{J}(P=0)$, $\Delta \mathcal{J}_0(P=0)$, and $\Delta \mathcal{J}_{00}(P=0)$ for $t$- and $u$-channels as well as  $\Delta \mathcal{J}(P=p_s)$, $\Delta \mathcal{J}_{0}(P=p_s)$, and $\Delta \mathcal{J}_{00}(P=p_s)$ for $s$-channels. The $\c{J}$ integrals/sums at $P=0$ can be shown to be related to $\mathcal{I}$, giving the volume difference:
\begin{align}
i\Delta\c{J}(0) &= -\frac{1}{4} \left[ 
\frac{1}{L^3}\sum_{\vec{q}} -\int\frac{d^3q}{(2\pi)^3} \right] 
\frac{1}{\omega_q^3} = \frac{d}{dm ^2} \left(i\Delta \mathcal{I}\right),\\
i\Delta \mathcal{J}_0(0) &= 0,
\end{align}
and
\begin{align}
i\Delta \c{J}_{00} (0) &= i\int \frac{dq_0}{2\pi} 
\left[ \frac{1}{L^3}\sum_{\vec{q}} -\int\frac{d^3q}{(2\pi)^3} \right] 
\left[\frac{1}{q^2-m ^2}+\frac{m^2}{(q^2-m ^2)^2}+\frac{\vec{q}^{\:2}}{(q^2-m ^2)^2}\right]\nn\\
&=i\Delta\c{I}+m^2 i\Delta\c{J}(0)+3\left(-\frac{1}{6} i\Delta\c{I}-\frac{1}{3}m^2 \frac{d}{dm^2} i\Delta\c{I}\right) \nn\\
&=\frac{1}{2} i\Delta \c{I}.
\end{align}

The power law volume dependence appears in the remaining integrals/sums. In those we keep $s$ away from the threshold value and take $\sqrt{s}=2\sqrt{k^2+m^2}$.  After performing the $q_0$ integral,  we separate the singular piece of the summand from the rest as
\beq
i\c{J}(p_s) = -\frac{1}{4L^3} \sum_{\vec{q}} \frac{1}{\omega_q} \frac{1}{\vec{q}^{\:2}-k^2}
= -\frac{1}{4\omega_k L^3} \sum_{\vec{q}}  \frac{1}{\vec{q}^{\:2}-k^2}
  +\frac{1}{4L^3} \sum_{\vec{q}} \frac{1}{\omega_q\omega_k} \frac{\omega_q-\omega_k}{\vec{q}^{\:2}-k^2}.
\eeq
The first term contains a singularity when the internal momentum coincides with the external momentum, while the second term is regular. The difference $\Delta \mathcal{J}(p_s)$ is then
\beq\label{eq:I}
	i\D \c{J}(p_s) = -\frac{1}{8\pi^2L\sqrt{s}} \c{S} \left( \frac{k^2L^2}{4\pi^2} \right)
				+\underbrace{\sum_{\vec{n}\neq 0} \int\frac{d^3q}{(2\pi)^3} 
					e^{iL\vec{q} \cdot \vec{n}}\, 
					\frac{\omega_q-\omega_k}{\vec{q}^{\:2}-k^2}
					\frac{1}{4\omega_k\omega_q}}_{i\Delta\mathcal{J}_{\rm exp}(p_s)} 
\eeq 
The first piece above is the promised universal function containing the power law volume dependence.   The summand in the second term contains only exponential finite volume corrections. This term, contributing to $\Delta \c{T}^{(1)}_{\rm exp}$, can be computed at the $s=4m^2, \vec{k}=0$ threshold point, 
\begin{align}\label{eq:Jexp}
	i\D \c{J}_{\rm exp}(p_s) 
		&= \frac{1}{16\pi^2} \frac{1}{L \sqrt{m^2 +k^2}}
			 \sum_{\vec{n}\neq 0} \frac{1}{|\vec{n}|} \int_{-\infty}^{\infty} d y
			 \frac{y\, {\rm Im} e^{i2\pi y |\vec{n}|}}
			 	{\sqrt{y^2+\frac{m^2L^2}{4\pi^2}} \left(\sqrt{y^2+\frac{m^2L^2}{4\pi^2}}
					+\sqrt{\frac{k^2L^2}{4\pi^2}+\frac{m^2L^2}{4\pi^2}}\right)} \nn \\
		&\simeq -\frac{1}{16\pi} \sum_{\vec{n}\neq 0} 
			\left[ K_0(\left| \vec{n} \right| m  L) \bar{L}_{-1}(\left| \vec{n} \right| m  L) 
  				+K_1(\left| \vec{n} \right| m  L) \bar{L}_{0}(\left| \vec{n} \right| m  L) 
  				-\frac{1}{\left| \vec{n} \right| m  L} \right],
\end{align}
where  $\bar{L}_\nu$ is the Struve function.  To get the second line of Eq.\eqref{eq:Jexp}, we have neglected terms which are suppressed by $\c{O}(k^2 / m^2)$ relative to the first.  For the two-pion system, this is approximately given by $\frac{k^2}{m^2} \simeq \frac{4 \pi |a|}{m^2 L^3} \ll 1$.  
The asymptotic expansion of $i\D\c{J}_{\rm exp}(p_s)$ is given by 
\begin{equation}
	i\D\c{J}_{\rm exp}(p_s) \simeq \frac{\sqrt{2\pi}}{(4\pi)^2} \frac{1}{(mL)^{3/2}}
				\sum_{n=|\vec{n}| \neq 0} 
					c(n)\, \frac{e^{-n\, m_\pi L}}{n^{3/2}} 
					\left[ 1 - \frac{5}{8} \frac{1}{n\, m_\pi L} + \ldots \right].
\end{equation}
Again, one can see that these volume corrections to the integral are exponentially suppressed.

The finite volume dependence of the other $s$-channel loop integral functions, 
$i\Delta\mathcal{J}_0(p_s)$ and $i\Delta\mathcal{J}_{00}(p_s)$ become simpler to evaluate by first observing that the summands can be separated into the following pieces:
\begin{align}
\frac{q_0}{q_0^2-\omega_q^2}\frac{1}{(q_0+p_{s0})^2-\omega_q^2} &=
\frac{1}{2 p_{s0}}\left[ \frac{1}{q_0^2-\omega_q^2}-\frac{1}{(p_{s0}+q_0)^2-\omega_q^2}
-\frac{(p_{s0})^2}{q_0^2-\omega_q^2}\frac{1}{(q_0+p_{s0})^2-\omega_q^2} \right], \nn \\
\frac{(q_0)^2}{q_0^2-\omega_q^2}\frac{1}{(q_0+p_{s0})^2-\omega_q^2} &=
\left( 1+\frac{\omega_q^2}{q_0^2-\omega_q^2} \right)\frac{1}{(q_0+p_{s0})^2-\omega_q^2}.\nn
\end{align}
One then obtains,
\begin{equation}\label{eq:I0}
i\D \mathcal{J}_0(p_s)
= -\frac{\sqrt{s}}{2} i\D \mathcal{J}(p_s)
\end{equation}
and
\begin{align}\label{eq:J00}
	i\D \c{J}_{00}(p_s) &= i\D\c{I} - \frac{1}{4L^3} \sum_{\vec{q}} \frac{1}{\omega_q} 
			\left( 1+\frac{\omega_k^2}{\vec{q}^{\:2}-k^2} \right) \nonumber\\
		&= \frac{1}{2} i\D \mathcal{I} + \frac{s}{4}\, i\D \mathcal{J}(p_s).
\end{align}

Having these tensor integrals/sums written in terms of the scalar integrals/sums $i\D\c{J}(p_s)$ and $i\D\c{I}$ and using the scattering amplitude in Eqs.~(\ref{eq:Tgeneral})
and~\eqref{eq:Ttensor}, one can now verify that the coefficient of $\mathcal{S}(\frac{k^2L^2}{4\pi^2})$ in the amplitude is what was promised in Eq.~(\ref{eq:Tgeneral}).  

%
%
%
%
%
\subsection{Exponential volume correction to the $I=2$ $\pi\pi$ correlator}

\begin{figure}
\includegraphics[width=0.6\textwidth]{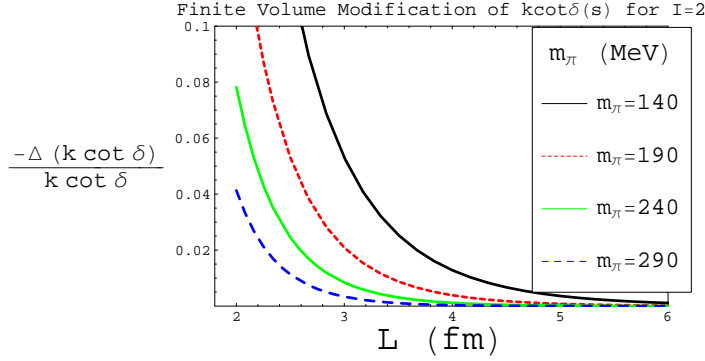} 
\caption{  Ratio of the magnitude of the exponential correction term $ \Delta (k\cot\delta) = -32\pi m_\pi\frac{\Delta T_{\rm exp}^{(1)}(4m_\pi^2)}{(T^{(0)}(4m_\pi^2))^2}$ to $k\cot\delta$ for different values of the pion mass.}
\label{fig:avsLP}
\end{figure}

Collecting the results for the sums/integrals in Eqs.~(\ref{eq:deltaI}-\ref{eq:J00}) and using the amplitude in Eq.~(\ref{eq:Ttensor}) we can now compute the correction term in Eq.~(\ref{eq:FVluscher}) for $I=2$ two-pion system near threshold. We find
\bea\label{eq:result}
\Delta (k\cot\delta(s)) &=& - 32\pi m_\pi\frac{\Delta T_{\rm exp}^{(1)}(4m_\pi^2)}{(T^{(0)}(4m_\pi^2))^2}
	    = \frac{8\pi}{m_\pi} \left[
		\frac{11}{3}i\Delta\mathcal{I}
		+ m_\pi^2 \frac{\partial}{\partial m_\pi^2} i\D\c{I} 
		+ 2   i\D\c{J}_{\rm exp}(4m_\pi^2) \right]\nn\\
	  &=& -\frac{m_\pi}{\sqrt{2\pi}} 
		\sum_{n=|\vec{n}| \neq 0} 
			c(n) \frac{e^{-n m_\pi L}}{\sqrt{n m_\pi L}} \left[
				1 -\frac{227}{24} \frac{1}{n m_\pi L} +\dots \right]. 
\eea
Equation~(\ref{eq:result}) is our main result (the first line being the exact one-loop answer and the second line the asymptotic expansion in $m_\pi L$).  In this expression, one can use either the bare parameters, the physical parameters or the finite volume parameters as the difference is higher order than we work in either the chiral expansion, or in the exponential dependence.  It is most convenient to use the values of $m_\pi(L)$ and $f_\pi(L)$ directly measured in a given lattice simulation.
%
In Fig.~\ref{fig:avsLP} we plot the ratio of $\Delta (k\cot\delta(s))$ to the one-loop value of $k\cot\delta(s)$  using Eq.~(\ref{eq:result}) as a function of $L$ for some reasonable values of $m_\pi$. 
We find the finite volume corrections to be relatively small, a few times smaller than the statistical and systematic errors quoted in recent simulations. An error of about $10\%$ was quoted in reference~\cite{Beane:2005rj} for the determination of the scattering length for a pion mass of $m_\pi \simeq 290\;{\rm MeV}$ and a box size of $L\simeq 2.5\;{\rm fm}$.%
\footnote{In Ref.~\cite{Beane:2005rj}, Beane \emph{et. al.} determined the $I=2$ $\pi\pi$ scattering length for various pion masses using a mixed action simulation with Domain-Wall valence quarks and staggered sea quarks~\cite{Bar:2005tu}.  Because of the mixed-action, the mesons composed of sea quarks and the mesons composed of valence quarks receive different mass shifts from the finite lattice spacing.  This means that even when the sea and valence quark masses are tuned equal, there are still partial quenching effects in the simulation.  In Ref.~\cite{Chen:2005ab}, the partial quenching and lattice spacing corrections to the $I=2$ $\pi\pi$ scattering length were worked out for this mixed action theory.  It was shown that the these two lattice artifacts were largely suppressed, and almost non-existent for the mass tunings used in Ref.~\cite{Beane:2005rj}.  However, as shown in Ref.~\cite{Chen:2005ab}, there are still partial quenching effects and in particular, in the $t$- and $u$-channel diagrams the hairpin contributions can be significantly more sensitive to the boundary effects.  For $I=2$ these effects are only exponential, and for the pion masses and box sizes used in Ref.~\cite{Beane:2005rj}, we have found they are the same order of magnitude as the corrections of this paper, and thus not-significant to the work of Beane \emph{et. al.}  These effects are being worked out in detail in Ref.~\cite{Sato-PQFV}}
 The finite volume correction from Eq.~(\ref{eq:result}) for these parameters is approximately $1\%$. These corrections however grow fast with the approach to the chiral limit, and they become non-negligible as smaller pion masses are used and statistical errors are reduced in simulations.

%
%
%
%
%
\section{Discussion}
We have described the leading exponential volume dependence expicitly for the scattering parameter of a $I=2$ two-pion system near threshold in a box, by extending the one-loop \CPT\ calculation of pion scattering~\cite{Gasser:1983yg} to include the volume dependence.  The exponentially suppressed volume corrections can distort the universal relation between the infinite volume scattering parameters and the power-law volume dependence of the two-particle system, especially as the chiral limit is approached.  An important point we want to stress is that the useful way to add the exponential volume dependence to the relation between infinite volume scattering parameters and the energy of the two-particle system in a box, is via Eq.~\eqref{eq:FVluscher}, which allows an understanding of the leading exponential volume dependence to $k\cot\delta(s)$.  This is contrast to the notion of studying the exponential volume dependence of the scattering length, the effective range etc., separately.

It is important to stress the limits of validity of the present calculation. On one hand, the pion masses should be small enough so that the chiral expansion is converging. From the experience acquired in the three flavor case, where kaon loops are a borderline case for the convergence of the expansion, one expects chiral perturbation theory to be useful for $m_\pi < 500 $ MeV (of course the exponential volume dependence for a $500$ MeV pion, or kaon will be negligible). Also, the box size has to be large enough so the usual power counting used here (the so-called ``$p$-counting") is valid. When $L$ is much smaller than the inverse pion mass, another power counting is required such as the $\epsilon$-~\cite{gasser_epsilon} or the $\epsilon^\prime$-regime~\cite{martin_epsilonprime}.  Additionally, we have neglected corrections which occur from higher loops, all of which are suppressed by additional factors of $(m_\pi / 4\pi f_\pi)^2$ and some of which are suppressed by additional exponential factors of $e^{-m_\pi L}$.  The diagrams with this extra exponential suppression result from two-loop diagrams where intermediate states in both loops are purely off-shell and hence ``going around the box".

We have focussed on the exponential corrections to phase shifts close to threshold.  One can easily extend this work to include the exponential volume dependence of the phase shifts at higher energies.  
Alternatively, one can access non zero momenta by using twisted or partially twisted boundary conditions to probe the low-momentum dependence of the scattering amplitude~\cite{bedaque_AB,deDivitiis:2004kq,Sachrajda:2004mi,Bedaque:2004ax,Tiburzi:2005hg,Mehen:2005fw,Flynn:2005in,Guadagnoli:2005be}.  This method may boost the entire two-particle system, however, requiring the extraction of scattering parameters in a boosted frame~\cite{Rummukainen:1995vs,Kim:2005gf,Christ:2005gi}. Our methods generalize trivially to this case also.  These methods can also be extended to other interesting two-hadron sytems~\cite{Beane:2003da,Beane:2003yx}, where these exponential volume effects may be more significant.

\acknowledgments
A.W-L. would like to thank David Lin for useful discussions.
This work was supported in part by the Director, Office of Energy Research,
Office of High Energy and Nuclear Physics, by the Office of Basic Energy
Sciences, Division of Nuclear Sciences, of the U.S. Department of Energy
under Contract No.~DE-AC03-76SF00098  and the National Science Council of ROC.
AWL is supported under DOE grant DE-FG03-97ER41014.

%
%
%
%
%
%
%
%
%
%
%
%


\end{document}